\pdfoutput=1

\documentclass[11pt,twoside,a4paper,cmspaper,final,collab]{cms-tdr}
\def\svnVersion{175956}\def\cmsCernNo{2012-248}\def\cmsCernDate{\today}\def\cmsMessage{Submitted to Physical Review D}

\begin{document}\cmsNoteHeader{FWD-10-004}

\hyphenation{had-ron-i-za-tion}
\hyphenation{cal-or-i-me-ter}
\hyphenation{de-vices}

\RCS$Revision: 159270 $
\RCS$HeadURL: svn+ssh://svn.cern.ch/reps/tdr2/papers/FWD-10-004/trunk/FWD-10-004.tex $
\RCS$Id: FWD-10-004.tex 159270 2012-11-27 16:34:28Z alverson $
\providecommand{\re}{\ensuremath{\cmsSymbolFace{e}}}
\providecommand{\wxi}{\ensuremath{\widetilde{\xi}}\xspace}
\providecommand{\wxim}{\ensuremath{\widetilde{\xi}^-}\xspace}
\providecommand{\wxip}{\ensuremath{\widetilde{\xi}^+}\xspace}
\providecommand{\wxipm}{\ensuremath{\widetilde{\xi}^\pm}\xspace}
\newlength\cmsFigWidth
\ifthenelse{\boolean{cms@external}}{\setlength\cmsFigWidth{0.98\columnwidth}}{\setlength\cmsFigWidth{0.6\textwidth}}
\ifthenelse{\boolean{cms@external}}{\providecommand{\cmsLeft}{top\xspace}}{\providecommand{\cmsLeft}{left\xspace}}
\ifthenelse{\boolean{cms@external}}{\providecommand{\cmsRight}{bottom\xspace}}{\providecommand{\cmsRight}{right\xspace}}
\ifthenelse{\boolean{cms@external}}{%
\newcommand{\scotchrule[1]}{\centering\begin{ruledtabular}\begin{tabular}{#1}}
\newcommand{\donescotchrule}{\end{tabular}\end{ruledtabular}}
}{
\newcommand{\scotchrule[1]}{\centering\begin{tabular}{#1}\hline\hline}
\newcommand{\donescotchrule}{\hline\hline\end{tabular}}
}
\cmsNoteHeader{FWD-10-004} 
\title{Observation of a diffractive contribution to dijet production in proton-proton collisions at \texorpdfstring{$\sqrt{s}=7\TeV$}{sqrt(s)=7 TeV}}

\date{\today}

\abstract{
The cross section for dijet production
in pp collisions at $\sqrt{s} = 7$\TeV
is presented as a function of  \wxi,
a variable that approximates the fractional momentum loss of the scattered proton
in single-diffractive events. The analysis is based on an
integrated luminosity of 2.7\nbinv collected with the CMS detector
at the LHC at low instantaneous luminosities,
and uses events with jet transverse momentum of at least 20\GeV.
The dijet cross section results are compared to the predictions of diffractive and
non-diffractive models. The low-\wxi data show a significant contribution from diffractive dijet
production, observed for the first time at the LHC. The associated rapidity gap survival
probability is estimated.
}

\hypersetup{%
pdfauthor={CMS Collaboration},%
pdftitle={Observation of a diffractive contribution to dijet production in proton-proton collisions at sqrt(s)=7 TeV},%
pdfsubject={CMS},%
pdfkeywords={CMS, physics, software, computing}}

\maketitle 

\section{Introduction}
A significant fraction of the total inelastic proton-proton cross section
at high energies is attributed to diffractive processes,
characterised by the presence of a large rapidity region $\Delta y$ with no hadrons,
usually called ``rapidity gap"
(rapidity is defined as $y=(1/2)\ln[(E+p_Z)/(E-p_Z)]$, where
$E$ and $p_Z$ are the energy and longitudinal momentum of the final-state
particle, respectively). Diffractive scattering is described in the framework of Regge
theory as mediated by a strongly interacting colour-singlet exchange with the vacuum
quantum numbers, the so-called ``pomeron trajectory''~\cite{regge}.
Diffractive events with a hard parton-parton
scattering are especially interesting because they can be studied in terms of
perturbative quantum chromodynamics (pQCD).
In diffractive events the proton emitting the pomeron either remains intact,
losing only a few per cent of its momentum,
or is found in a low mass excited state.
In addition, since the vacuum quantum numbers are exchanged, no particles
are produced in a large rapidity range adjacent to the scattered proton
(or its dissociation products).

Diffraction with a hard scale has been studied
in  proton-antiproton (\Pp\Pap) and electron-proton (\Pe\Pp) collisions at
CERN~\cite{cern_diffraction},
Fermilab~\cite{tevatron_diffraction1, tevatron_diffraction2,cdf_dijet,newcdf}, and
DESY~\cite{hera_diffraction1,hera_diffraction2,zeuspdf,h1phpjet}.
Such hard diffractive processes can be described
in terms of the convolution of diffractive parton distribution functions (dPDFs) and
hard scattering cross sections, which are calculable in pQCD.
In this approach, the pomeron is
treated as a colour-singlet combination of partons
with the vacuum quantum numbers.
The dPDFs have been determined by the HERA
experiments~\cite{hera_diffraction1,zeuspdf} by means of QCD fits to
inclusive diffractive deep inelastic scattering data, and have been
successfully used to describe different
hard diffractive processes in \Pe\Pp\ collisions. This success is based on
the factorisation theorem
for diffractive electron-proton interactions, and on the validity of the QCD
evolution equations for the dPDFs~\cite{fracture1,collins,fracture2}.
However, in hard diffractive hadron-hadron collisions factorisation
does not hold because of soft scatterings between the spectator partons,
leading to the suppression of the observed diffractive cross section.
The suppression is quantified by the so-called ``rapidity gap survival probability''~\cite{survival},
which is a non-perturbative quantity with large theoretical uncertainties~\cite{svp1,svp2,svp3,svp4}.
It was measured to be about $10\%$
in diffractive dijet production in \Pp\Pap\ collisions at the Tevatron~\cite{cdf_dijet}.

This paper presents a study of dijet production in proton-proton collisions
at a centre-of-mass energy of $\sqrt{s}=7$\TeV. The data were collected with the
Compact Muon Solenoid (CMS) detector at the Large Hadron Collider (LHC)
in 2010 and correspond to an integrated luminosity of 2.7\nbinv.
The cross section for production of dijets
is presented as a function of \wxi,
a variable that approximates the fractional momentum loss of the proton,
for events in which both jets have transverse momenta $p_{\mathrm{T}}^\mathrm{j1,j2}>20$\GeV and
jet axes in the pseudorapidity range $|\eta^\mathrm{j1,j2}|<4.4$.
Pseudorapidity is
defined as $\eta = -\ln[\tan(\theta/2)]$, where $\theta$ is the polar angle
relative to the anticlockwise proton beam direction, and
is equal to the rapidity in the limit of a massless particle.
The measurements are compared to the predictions of non-diffractive and
diffractive models, and the rapidity gap survival probability is estimated.

The paper is organised as follows: in Section~\ref{sec:setup} a brief description
of the CMS detector is provided. The definitions of the kinematic variables
are introduced in Section~\ref{sec:kine}. The event selection is explained in Section~\ref{sec:selection}.
Section~\ref{sec:MC} describes the main features of the Monte Carlo (MC) generators used
in this analysis. The cross section determination for dijets
as a function of \wxi and the systematic uncertainties
of the measurements are discussed in Section~\ref{sec:xsec}. The results are presented
in Section~\ref{sec:results}, and the summary is given in Section~\ref{sec:summary}.

\section{Experimental setup\label{sec:setup}}

A detailed description of the CMS detector can be found elsewhere~\cite{cms}.
The  central feature of the CMS apparatus is a superconducting solenoid, of 6\unit{m}
internal diameter. Within the field volume are the silicon pixel and strip tracker,
the crystal electromagnetic calorimeter (ECAL) and the brass-scintillator hadronic
calorimeter (HCAL).
The tracker measures charged particles within the pseudorapidity
range $\abs{\eta}< 2.4$. ECAL and HCAL provide coverage in pseudorapidity
up to $|\eta|<3$ in the barrel region and two endcap regions.
The HCAL, when combined with the ECAL, measures jets with an energy resolution
$\Delta  E/E  \approx 100\%/\sqrt{E~(\GeVns)} \oplus 5\%$.
The calorimeter cells are grouped in projective towers, of granularity
$\Delta\eta \times \Delta\phi = 0.087 \times 0.087$
at central rapidities and $0.175 \times 0.175$ at forward rapidities,
where $\phi$ is the azimuthal angle in radians. In addition to the barrel
and endcap detectors, CMS has extensive forward calorimetry.
The forward part of the hadron calorimeter, HF, consists of steel absorbers
and embedded radiation-hard quartz fibers, which provide a fast collection
of Cherenkov light.
The pseudorapidity coverage of the HF is $2.9 < |\eta | < 5.2$.
In the current analysis only the range $3.0 < |\eta | < 4.9$ was used,
thus restricting the data to a region of well understood reconstruction
efficiency.
The first level of the CMS trigger system, composed of custom hardware
processors, uses information from the calorimeters and muon detectors to
select the most interesting events in a fixed time interval of less than $4\mus$.
The High Level Trigger processor farm further decreases the event rate
from around $100\unit{kHz}$ to around $300\unit{Hz}$, before data storage.

\section{Kinematics and cross sections\label{sec:kine}}
\begin{figure}[hbtp]
  \begin{center}
    \includegraphics[width=\cmsFigWidth]{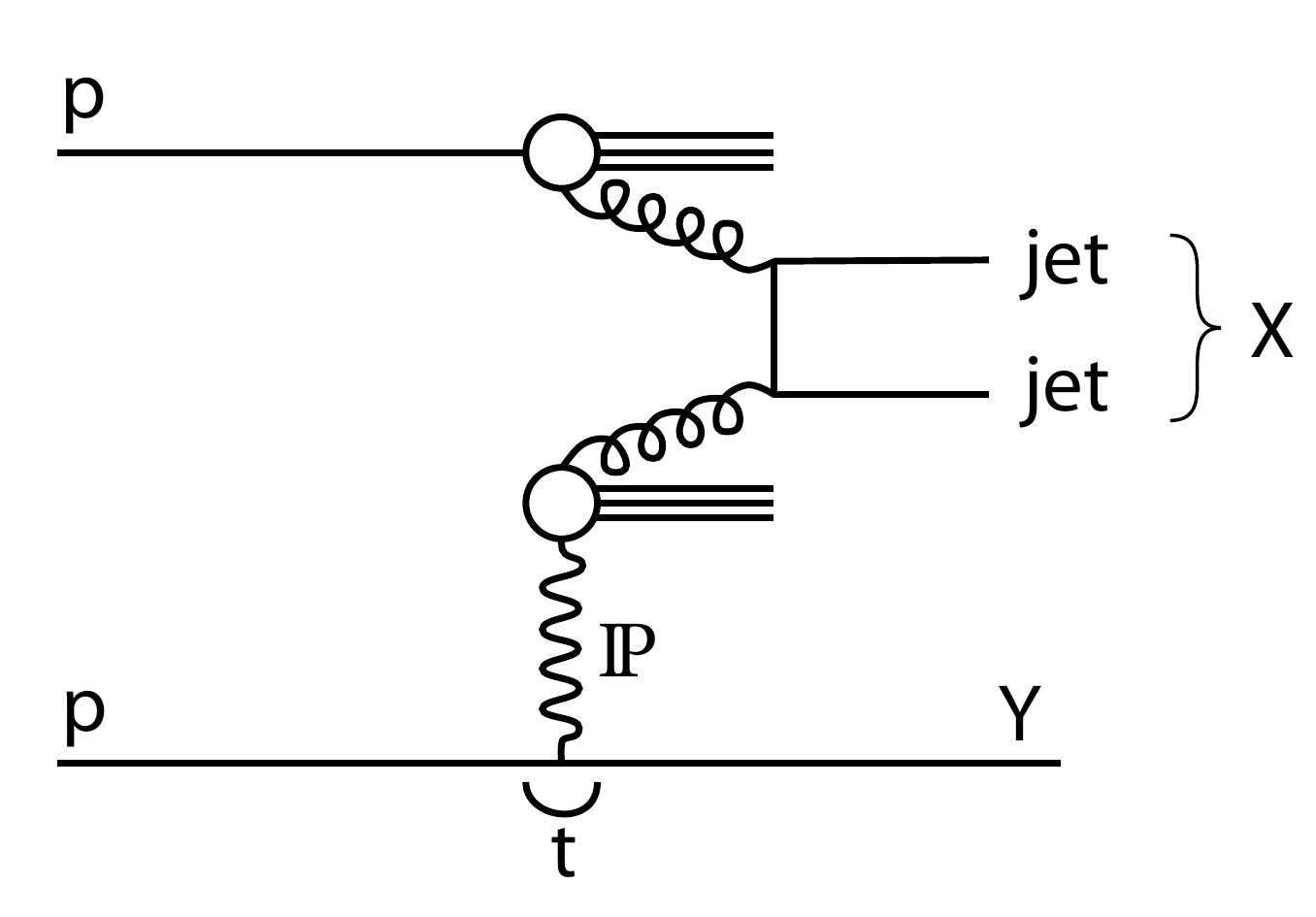}
     \caption{Schematic diagram of diffractive dijet production.
The diagram shows the example of the $\Pg\Pg \to$ jet jet process; the
$\Pq\Pq$ and $\Pg\Pq$ initial states  also contribute.
}
    \label{fig:dijet_graph}
  \end{center}
\end{figure}

Diffractive dijet production (Fig.~\ref{fig:dijet_graph})  is characterised by the presence of a high-momentum proton
(or a system $Y$ with the same quantum numbers as the proton) with fractional momentum
loss smaller than a few per cent
and a system $X$, which contains high-\pt jets and is separated from the
proton by a large rapidity gap, with $\Delta y \geq 3$ or 4 units.
The kinematics of this reaction is described by the masses
of the systems $X$ and $Y$, $M_{X}$ and $M_{Y}$,
and the squared four-momentum transfer $t$ at the proton vertex.
For the events selected in this analysis both $M_{X}$ and $M_{Y}$
are much smaller than $\sqrt{s}$.

The cross section for single-diffractive (SD) dijet production
(\ie when the forward-going system $Y$ is a proton)
is usually expressed in terms of the variable $\xi=M_{X}^{2}/s$,
which approximates the fractional momentum loss of the scattered proton.
Under the assumption of QCD factorisation, the cross section can be written as

\begin{equation}
\frac{\rd\sigma}{\rd\xi\, \rd t}=\sum \int \rd x_{1} \,\rd x_{2}\, \rd\hat{t}\, f(\xi,t) f_{\mathbb{P}}(x_{1},\mu) f_{p}(x_{2},\mu) \frac{\rd\hat{\sigma}(\hat{s},\hat{t})}{\rd\hat{t}},
\end{equation}

where the sum is over all parton flavours.
The variables $x_{1,2}$ are the parton momentum fractions in the pomeron
and proton, the scale at which the PDFs are evaluated is indicated with
$\mu$, and $\hat{\sigma}(\hat{s},\hat{t})$
is the hard-scattering subprocess cross section, which is a function of
the partonic centre-of-mass energy  squared $\hat{s}$ and
momentum transfer squared $\hat{t}$.
The function $f_{p}(x_{2},\mu)$  is the inclusive PDF of the proton that breaks up,
while the dPDF of the surviving proton is written as
$f_\text{diff}(\xi,t,x_{1},\mu)=f(\xi,t) f_{\mathbb{P}}(x_{1},\mu)$,
where $f(\xi,t)$ is the so-called pomeron flux and $f_{\mathbb{P}}(x_{1},\mu)$
is the pomeron structure function.
The cross section dependence on $\xi$ and $t$ is driven by the pomeron flux,
usually parameterised according to Regge theory as

\begin{equation}
f(\xi,t) = \frac{\re^{Bt}}{\xi^{2\alpha_{\mathbb{P}}(t)-1}},
\end{equation}

where $\alpha_{\mathbb{P}}(t)$
is the pomeron trajectory and $B$ is the slope parameter.
This ansatz is consistent with the HERA ep data~\cite{hera_diffraction1,hera_diffraction2,zeuspdf},
but is known not to hold between the ep and the Tevatron ($\Pp\Pap$)
data~\cite{tevatron_diffraction1, tevatron_diffraction2,cdf_dijet,newcdf},
where an extra suppression (gap survival probability) factor is needed.

\begin{figure}[hbtp]
  \begin{center}
    \includegraphics[width=0.48\textwidth]{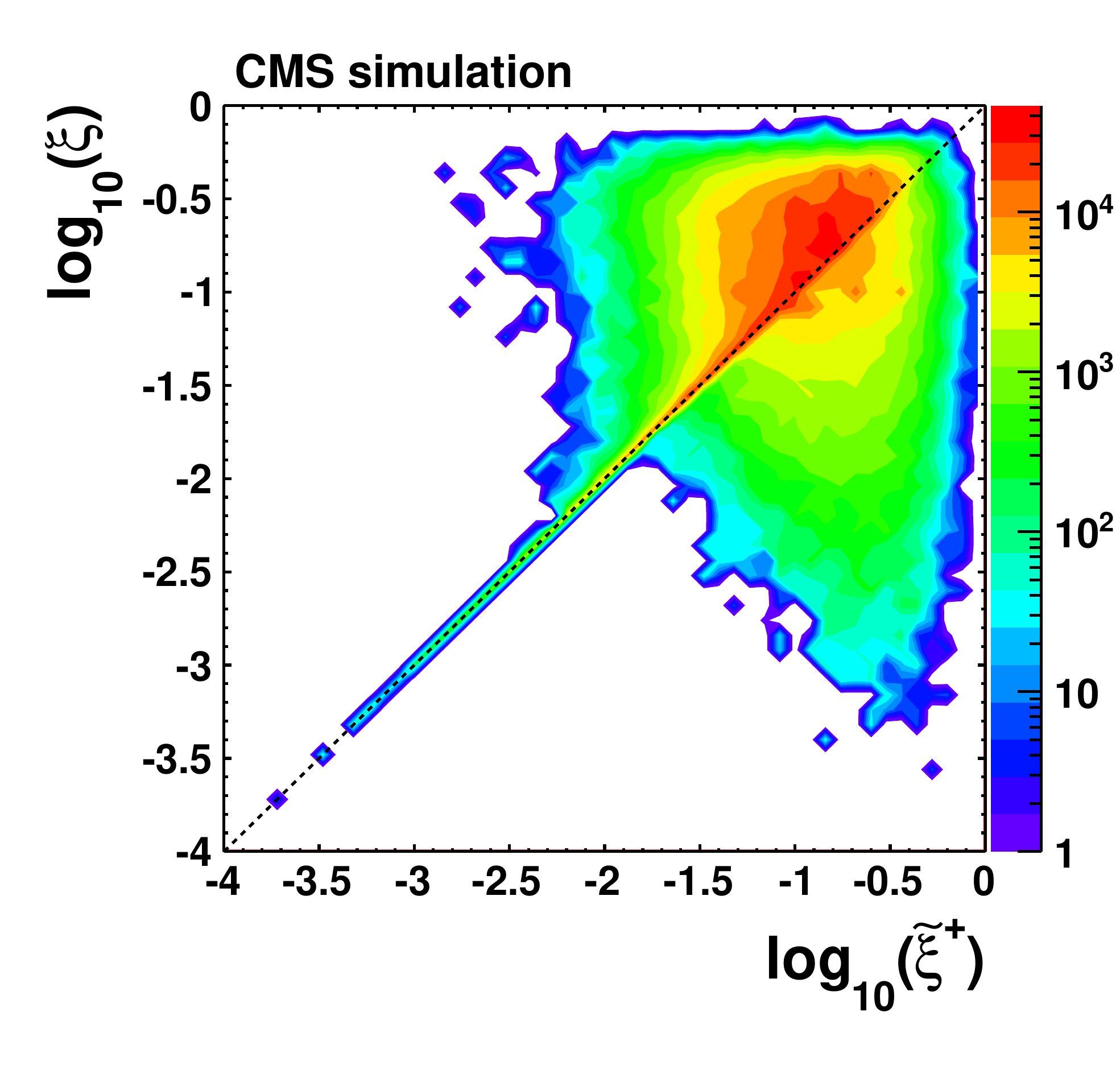} \includegraphics[width=0.48\textwidth]{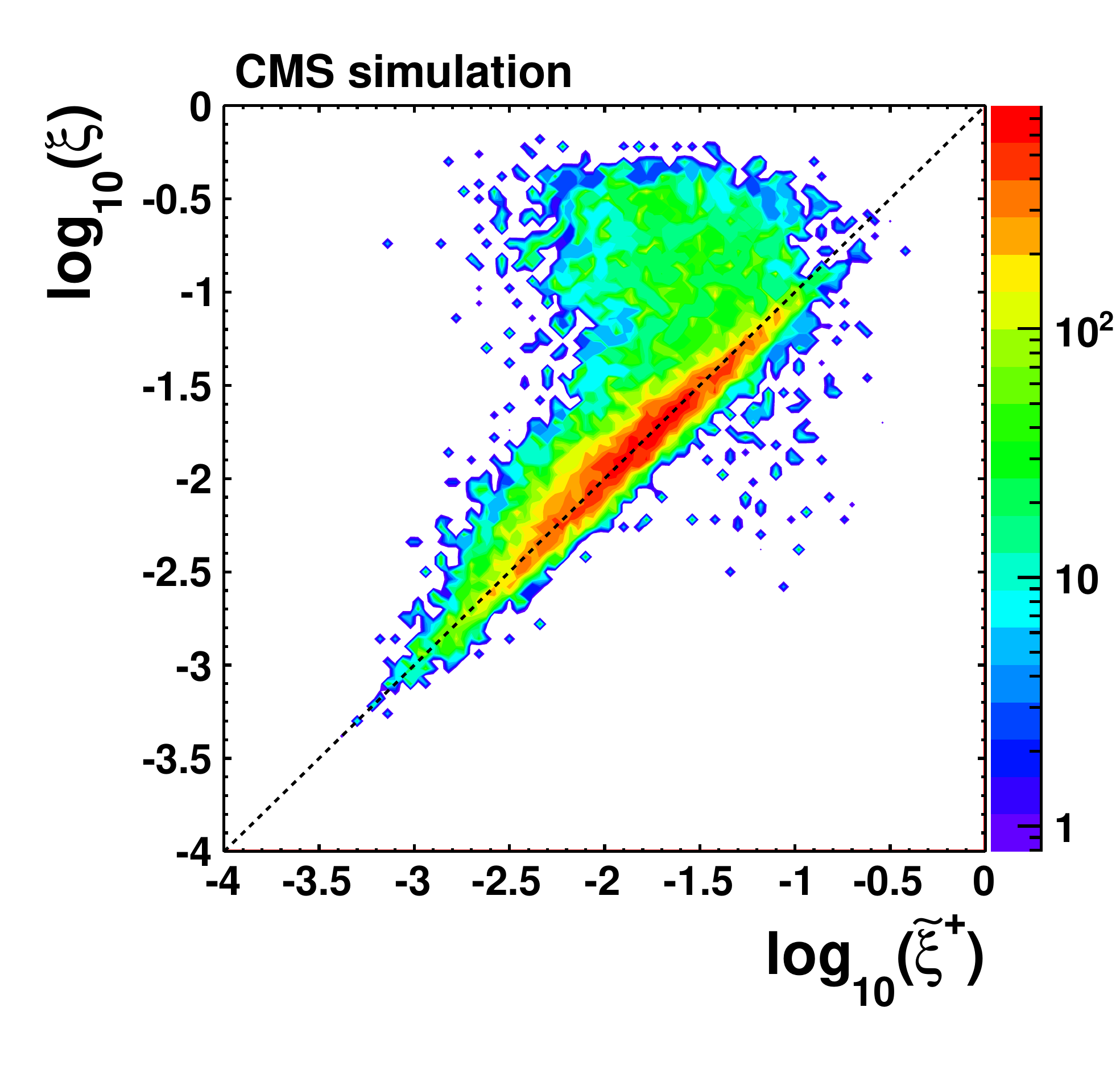}
     \caption{
The generated $\xi$ versus generated (\cmsLeft) and reconstructed (\cmsRight) $\wxip$ correlations
for single-diffractive dijet events simulated by \PYTHIA{}8; events in the \cmsRight panel are those passing the selection described in Section~\ref{sec:selection}.
}
    \label{fig:xi_correlation}
  \end{center}
\end{figure}

In this analysis $\xi$ is approximated by the variables
$\wxip$ (system $X$ going in the -$z$ direction)
and $\wxim$ (system $X$ going in the +$z$ direction)
defined at the level of stable particles as

\begin{equation}
\wxipm= \frac{\sum{(E^{i} \pm p_{z}^{i}})}{\sqrt{s}} ,
\end{equation}

where $E^{i}$ and $p_{z}^{i}$ are the energy and longitudinal momentum of the $i^{th}$
final-state particle with
$-\infty<\eta<4.9$ for $\wxip$ and $-4.9<\eta<+\infty$ for
$\wxim$.
In the region of low  $\wxipm$, this variable is a good approximation
of  $\xi$ for single-diffractive events.
This is illustrated
for single-diffractive dijet events simulated by \PYTHIA{}8~\cite{pythia8}
in Fig.~\ref{fig:xi_correlation}, where the correlations
between
the values of $\xi$  and $\wxip$, determined at generated and reconstructed (see Section~\ref{sec:selection}) are shown. The mass of the
forward-going system $Y$, which includes all particles with
$\eta>4.9$ (or $\eta<-4.9$),
was also estimated with the \PYTHIA{}8 generator;
the mass is limited by the pseudorapidity range and is typically smaller
than 30--40\GeV, with average $\sim$5\GeV.

\section{Event selection\label{sec:selection}}

The data were collected with the CMS detector in 2010 at low luminosities.
The average number of extra pp interactions for any given event
(the so called pile-up interactions) in the data is 0.09.
The low number of pile-up interactions simplified the extraction of the
diffractive signal, since the particles produced in such
interactions may fill the rapidity gap and hence reduce the
visible diffractive cross section. However, the requirement of low pile-up
limits the available data sample since only a small amount  of low-luminosity runs
was collected.

At the trigger level events were selected by requiring at least one jet with
uncorrected transverse momentum greater than 6\GeV.
The efficiency of the trigger, estimated using a minimum-bias data sample,
was found to be greater than $95\%$  for the dijet events considered in
this analysis.

Offline, the jets were reconstructed with the anti-$k_{T}$ inclusive jet finding
algorithm~\cite{anti-kt} with distance parameter of 0.5.
The jet clustering
algorithm was used to reconstruct jets from particle-flow (PF) objects~\cite{PF1},
which are particle candidates obtained by combining the information
of the tracking system and of the calorimeters in an optimal way.
The reconstructed jet momenta were fully corrected to the level of stable
particles (with lifetime $\tau$ such that $c\tau >10$ mm,
hereafter referred to as ``particle level"), by means of a procedure
partially based on MC simulation and partially
on data~\cite{jet_correction}.

The quantities $\wxip$ and $\wxim$ were
reconstructed using Eq.~(3) from the energies and longitudinal momenta
of all PF objects measured in the $|\eta|<4.9$ range.
For charged PF objects ($|\eta|<2.4$, the region covered by the tracker)
a minimum transverse momentum of 0.2\GeV was required.
In the forward region, $3.0<|\eta|<4.9$, particularly relevant for
this analysis, PF candidates were selected
with energy greater than 4\GeV.
A constant scale factor $C=1.45\pm0.04$, determined from the MC simulation
by comparing the generated and reconstructed values of  $\wxipm$,
is applied to the measured $\wxipm$. The error on the
correction factor $C$ is estimated by changing the MC models used to
evaluate it.
The value of $C$ reflects the fact that not all final-state particles are detected
because of the limited acceptance and imperfect response of the detector.
It also takes into account the inefficiency
of PF object reconstruction.
In practice, $C$ acts as a scale calibration for $\wxipm$;
it depends only slightly on the value of $\wxipm$ and
on the MC generator used. This dependence, of the order of a few per cent,
is included in the systematic uncertainty.
The resolution of $\wxipm$, in the region of the present
measurement, is ${\sim}25\%$, and practically independent of $\wxipm$.

Events were selected offline by
applying the following requirements:

\begin{itemize}

\item
the jets should pass the standard CMS quality criteria~\cite{jet_correction};

\item
events should have at least two jets, each with transverse
momentum, corrected to particle level,
greater than 20\GeV. This requirement ensures high trigger efficiency;

\item
the axes of the two leading jets (jets were ordered in \pt with the
first, leading jet having the highest \pt)
should be in the pseudorapidity region $|\eta^\mathrm{j1,j2}|<4.4$
so that the reconstructed jets are fully contained in the detector;

\item
a primary vertex should be within a longitudinal distance $|z|<24$ cm of the centre of CMS;

\item
beam-scraping events,
in which long horizontal sections of the pixel tracker are hit
by charged particles travelling parallel to the beam, were rejected
with a special algorithm~\cite{scraping};

\item
to enhance the diffractive contribution, the requirements $\eta_\text{max}<3$
($\eta_\text{min}>-3$) were also applied. Here  $\eta_\text{max}$ ($\eta_\text{min}$)
is the pseudorapidity of the most forward (backward) PF object.
The $\eta_\text{max}$ ($\eta_\text{min}$) selection together with the pseudorapidity
coverage of the detector, $|\eta|<4.9$, is equivalent to imposing a pseudorapidity
gap of at least 1.9 units, with no PF objects with energy greater than 4\GeV in the HF calorimeter.

\end{itemize}

The number of selected events before the $\eta_\text{max}$  ($\eta_\text{min}$)
requirement is 277\,953.  The number of events passing also the $\eta_\text{max}<3$
($\eta_\text{min}>-3$) selection is 804 (774); of these, 222 (220) have
$\wxip<0.01$ ($\wxim<0.01$).
The differential cross section for dijet production was calculated
separately as a function of $\wxip$ and $\wxim$.
The final results were averaged, and the average is presented as a function of \wxi.

The $\eta_\text{max}$, $\eta_\text{min}$ requirements reject
most pile-up interactions. The remaining pile-up background was estimated with
minimum-bias MC samples (\PYTHIA{}6 Z1 and \PYTHIA{}8,
see next Section)
and was found to be less than $2\%$.

\section{Monte Carlo simulation\label{sec:MC}}
The simulation of non-diffractive (ND) dijet events was performed with the
\PYTHIA{}6 (version 6.422)~\cite{pythia6}
and \PYTHIA{}8 (version 8.135)~\cite{pythia8} generators;
the events were generated in \PYTHIA{}6 with tunes Z2~\cite{z2}
and D6T~\cite{D6T}, and in \PYTHIA{}8 with tune 1~\cite{pythia8}.
The more recent \PYTHIA{}8 tune 4C~\cite{p84c} yields similar results as the
tune 1 used here.
Minimum-bias events were generated with \PYTHIA{}6 tune Z1~\cite{z2} and
with \PYTHIA{}8 tune 1.

Diffractive dijet events were simulated with the \textsc{pompyt}~\cite{pompyt},
\textsc{pomwig}~\cite{pomwig}, and \PYTHIA{}8 generators.
The \PYTHIA{}8 generator can simulate inclusive, non-diffractive as well as
diffractive dijet events; separate samples were produced for the two processes.
The modelling of diffractive events in these generators is based on the
Ingelman and Schlein approach~\cite{ingelman},
which considers the diffractive reaction as a two-step process:
one proton emits a pomeron with fractional momentum $\xi$ and then the pomeron
interacts with the other proton.
All three diffractive generators were used with dPDFs from the same fit to
diffractive deep inelastic scattering data (H1 fit B~\cite{hera_diffraction1}).
The parameterisation of the pomeron flux in \textsc{pompyt} and \textsc{pomwig}
is also based on the QCD fits to the HERA data~\cite{hera_diffraction1},
while it is different in \PYTHIA{}8~\cite{diffractioninpythia}.
This leads to different predictions
for the diffractive cross sections calculated by \PYTHIA{}8 and
\textsc{pompyt} or \textsc{pomwig} (notably in their normalisation).
The effect of the rapidity gap survival probability is not simulated
in any of the three diffractive generators.

The main difference between \textsc{pompyt} and \textsc{pomwig} is that
\textsc{pompyt} uses the \PYTHIA{} framework while \textsc{pomwig} is
based on \textsc{herwig}~\cite{herwig}.
Both programmes generate  single-diffractive dissociation.
In \PYTHIA{}8 double-diffractive dissociation (DD), in which both protons
dissociate, is also included.
The contribution from central diffractive dissociation, in which both protons stay intact,
was estimated with  \textsc{pomwig}.
It amounts to ${\sim}1\%$ of the diffractive contribution in the
\wxi region used in the analysis and was neglected.
Only pomeron exchange was assumed; the Reggeon exchange contribution
in the region $\wxi<0.01$ was estimated with  \textsc{pompyt}
and was found to be less than 2\%, and less than 1\% in the lowest
\wxi bin used in the analysis.

The diffractive component of the dijet cross section was also computed at
next-to-leading (NLO) accuracy with the \POWHEG~\cite{powheg} framework
using the CTEQ6M PDF for the proton that breaks up and H1 fit B for the
dPDF. The parton shower and hadronisation were carried out with
\PYTHIA{}8 (tune 1).

\begin{table*}[htbH]
\begin{center}
\topcaption{
Monte Carlo generators used in this work with details on their model ingredients.
\label{tab:page_layout2}}
\scotchrule[lllll]

Model   & PDF & dPDF    &  Parameter tune & Process \\
\hline
\PYTHIA{}6 & CTEQ6L1 & none &  Z2, D6T & Non-diffractive jets   \\
\PYTHIA{}8 &  CTEQ5L   & H1 fit B & Tune 1 & Diffractive plus non-diffractive jets  \\
\textsc{pompyt} &  CTEQ6L1   &  H1 fit B &  \PYTHIA{}6 D6T & Diffractive jets only  \\
\textsc{pomwig} &  CTEQ6L1   &  H1 fit B &  \textsc{herwig}  & Diffractive jets only \\
\POWHEG &  CTEQ6M   &  H1 fit B &   \PYTHIA{}8 tune1 & Diffractive jets only \\
\donescotchrule
\label{tab:mcs}
\end{center}
\end{table*}

The generators used are listed in Table~\ref{tab:mcs} along
with some of their features.
All generated events were processed through the simulation
of the CMS detector, based on \GEANTfour~\cite{geant4} and reconstructed
in the same manner as the data. All samples were generated without pile-up.
The measurements were corrected for detector acceptance and resolution with a
suitable combination of non-diffractive (\PYTHIA{}6 Z2) and diffractive
(\textsc{pompyt}) models (see Section\ref{ssec:reconstructed}).

\begin{figure*}[hbtp]
  \begin{center}
    \includegraphics[width=0.48\textwidth]{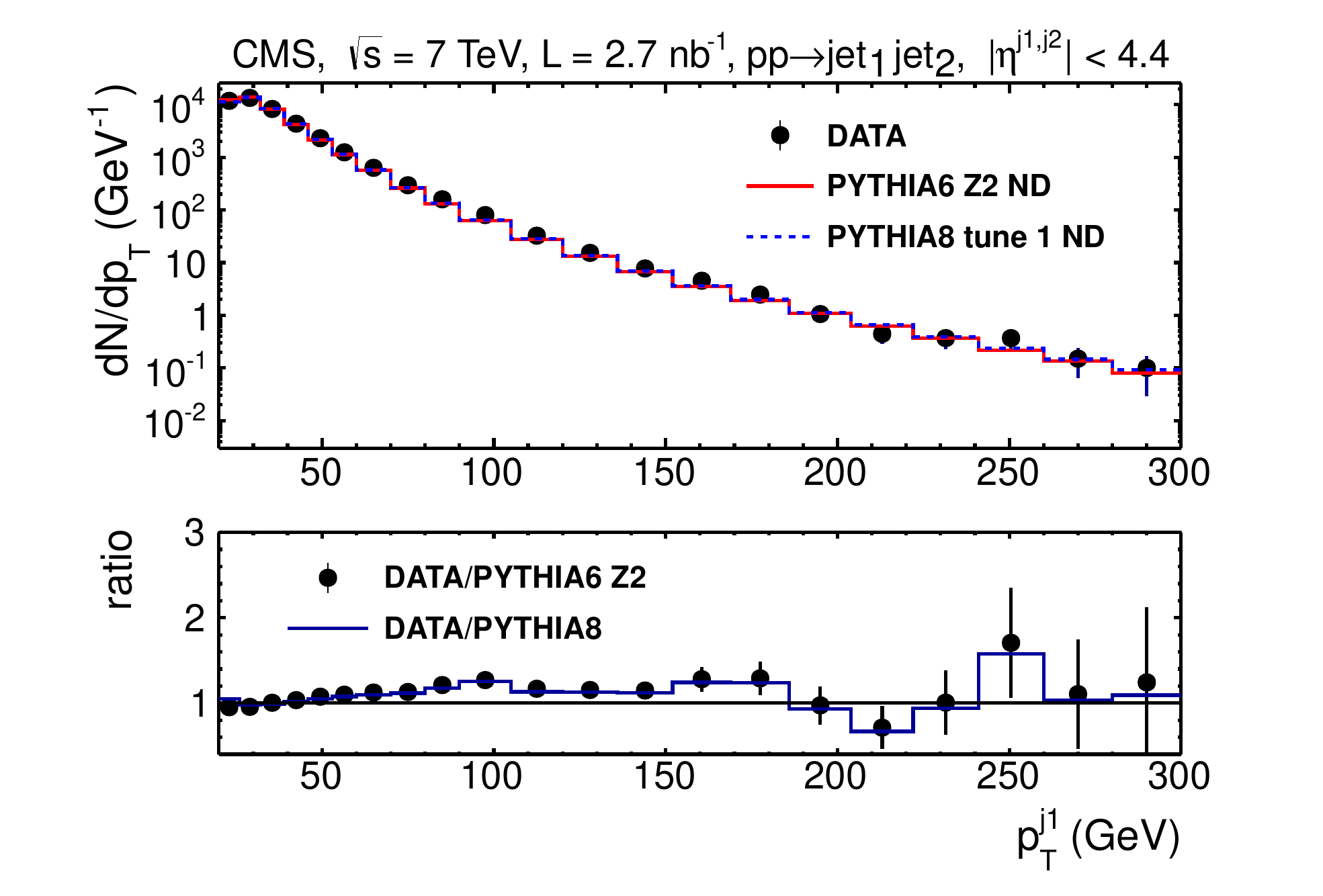}
    \includegraphics[width=0.48\textwidth]{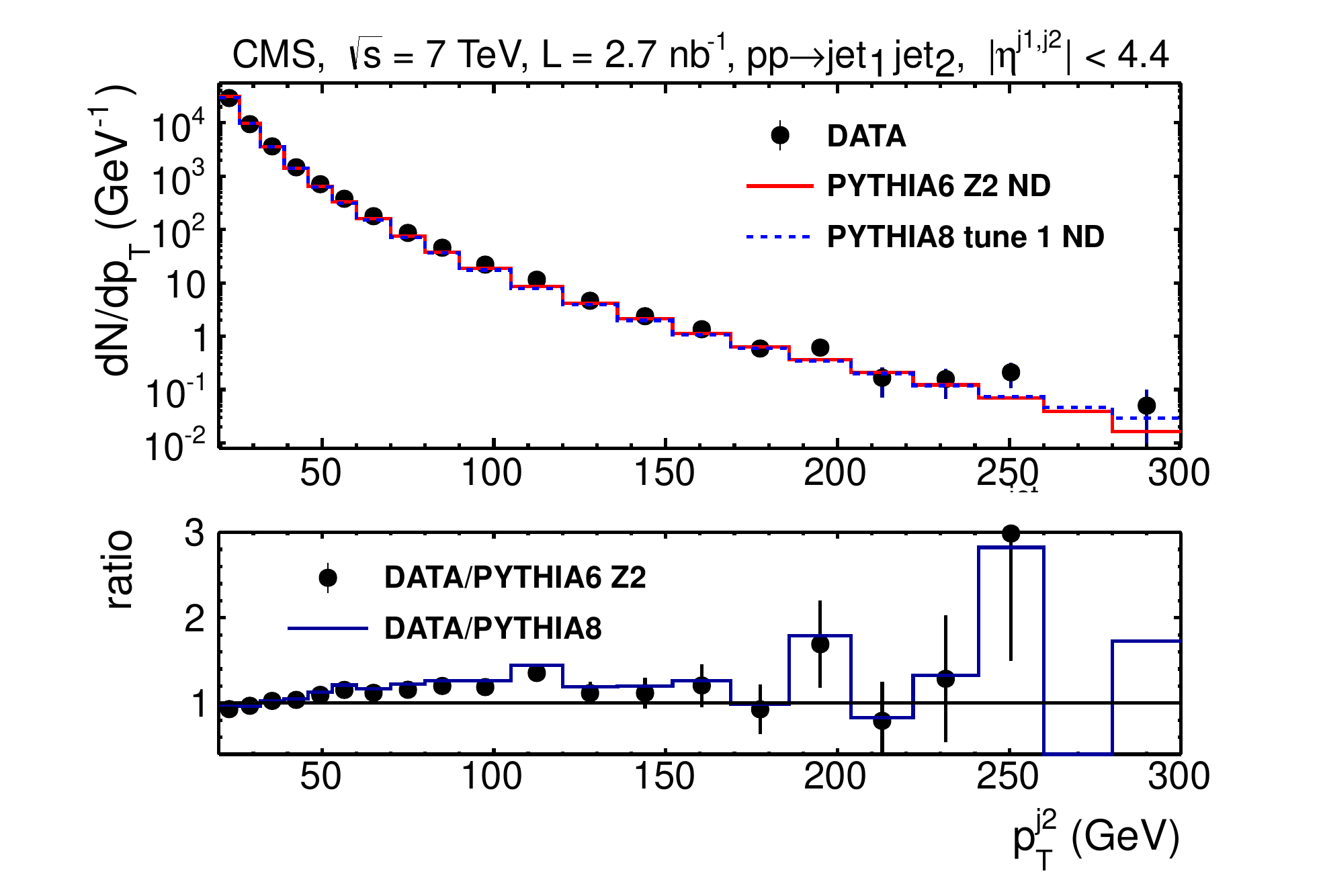}
    \caption{
Reconstructed transverse-momentum distributions of the leading (left)
and second-leading (right) jets
(black dots) compared to
detector-level MC simulations (histograms) generated with two non-diffractive
models (\PYTHIA{}6 Z2 and
\PYTHIA{}8  tune 1). The error bars indicate the statistical
uncertainty. The MC distributions are normalised to the
number of events in the corresponding distributions for the data.
The ratios of the data and MC distributions are also shown.
}
    \label{fig:jet_pt}
  \end{center}
\end{figure*}

Figure~\ref{fig:jet_pt} shows the comparison between  the uncorrected data
and detector-level MC simulations
for the reconstructed $\pt$ distributions of
the leading and second-leading jets with axes in the range
$|\eta^\mathrm{j1,j2}|<4.4$.
The simulated distributions are normalised to the
number of events in the corresponding distributions
for the data. The data and MC simulations are in agreement,
for both \PYTHIA{}6 Z2 and \PYTHIA{}8 tune 1.

\begin{figure}[hbtp]
  \begin{center}
    \includegraphics[width=\cmsFigWidth]{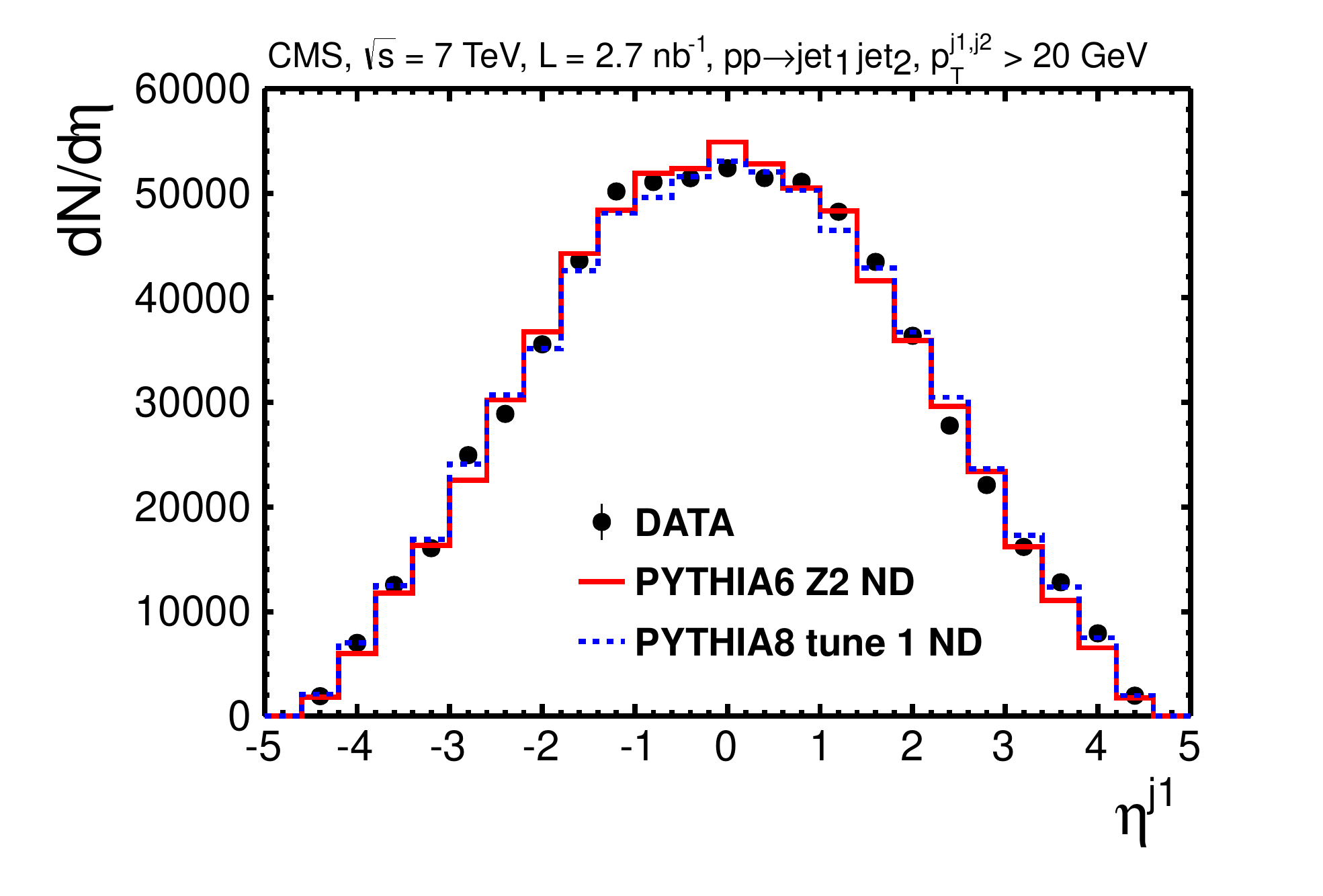}
    \includegraphics[width=\cmsFigWidth]{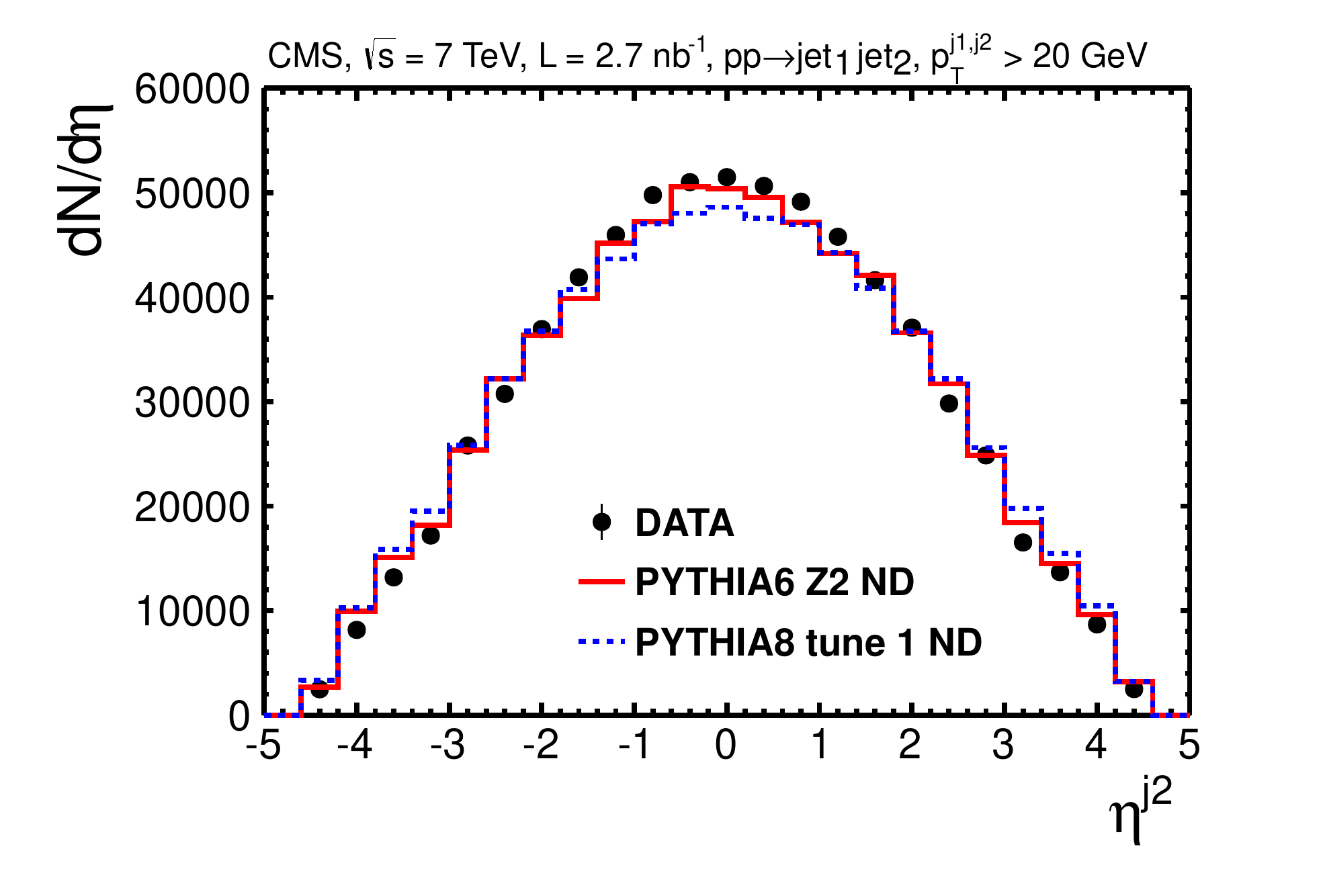}
    \caption{
Reconstructed pseudorapidity distributions of the leading (top) and second leading
(bottom) jets
(black dots) compared to detector-level MC
simulations (histograms) generated with two non-diffractive models
(\PYTHIA{}6 Z2  and \PYTHIA{}8  tune 1).
The statistical uncertainties are smaller than the data points.
The MC distributions are normalised to the number of events in
the corresponding distributions for the data.
}
    \label{fig:jet_eta}
  \end{center}
\end{figure}

Figure~\ref{fig:jet_eta} presents the comparison between data and MC simulations
for the reconstructed (detector-level)  pseudorapidity
distributions of the leading and second-leading jets.
Also here, the MC distributions are normalised to the number
of events in the data.
Data are better described by \PYTHIA{}6 tune Z2  than by \PYTHIA{}8
tune 1.

\begin{figure}[hbtp]
  \begin{center}
    \includegraphics[width=\cmsFigWidth]{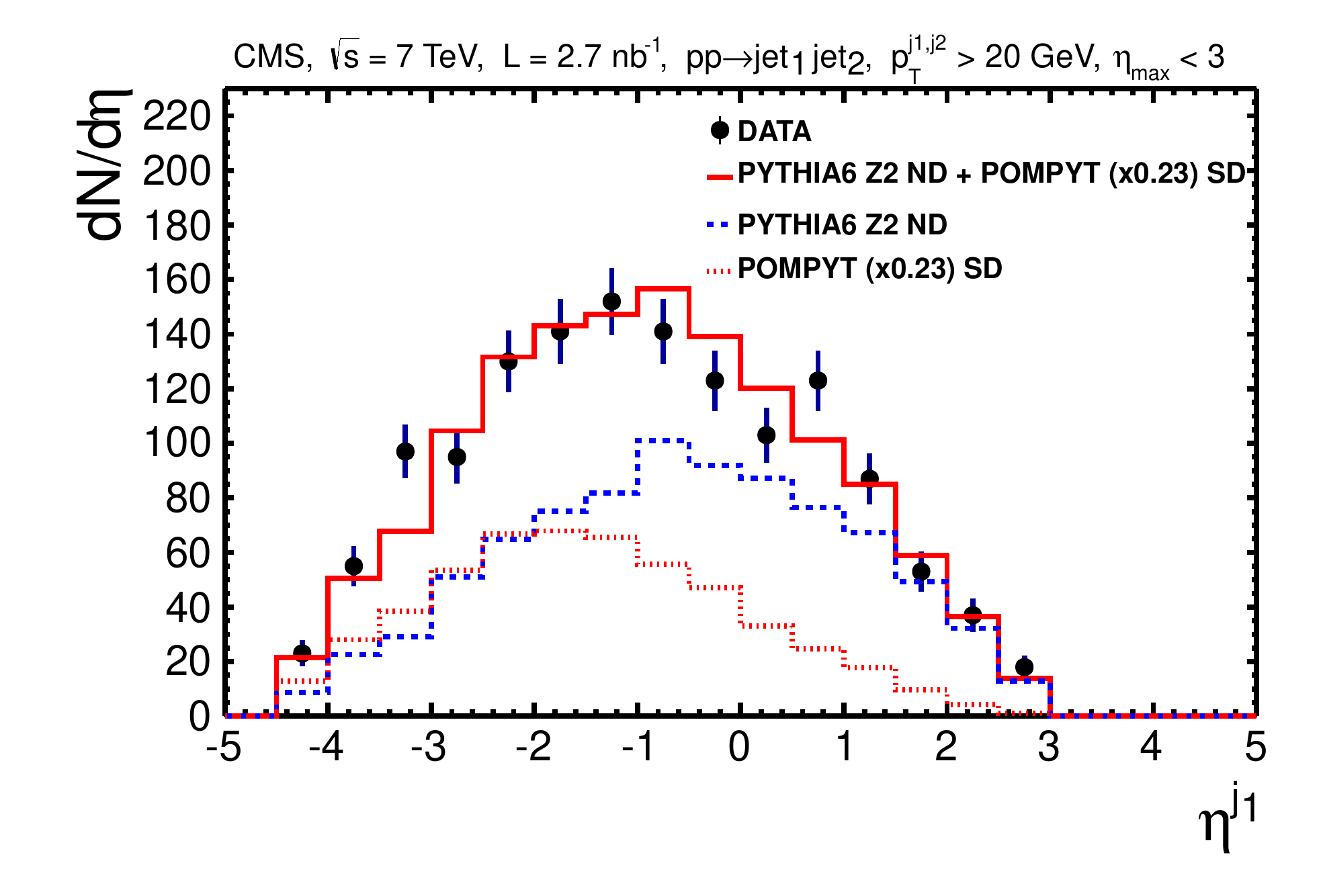}
    \includegraphics[width=\cmsFigWidth]{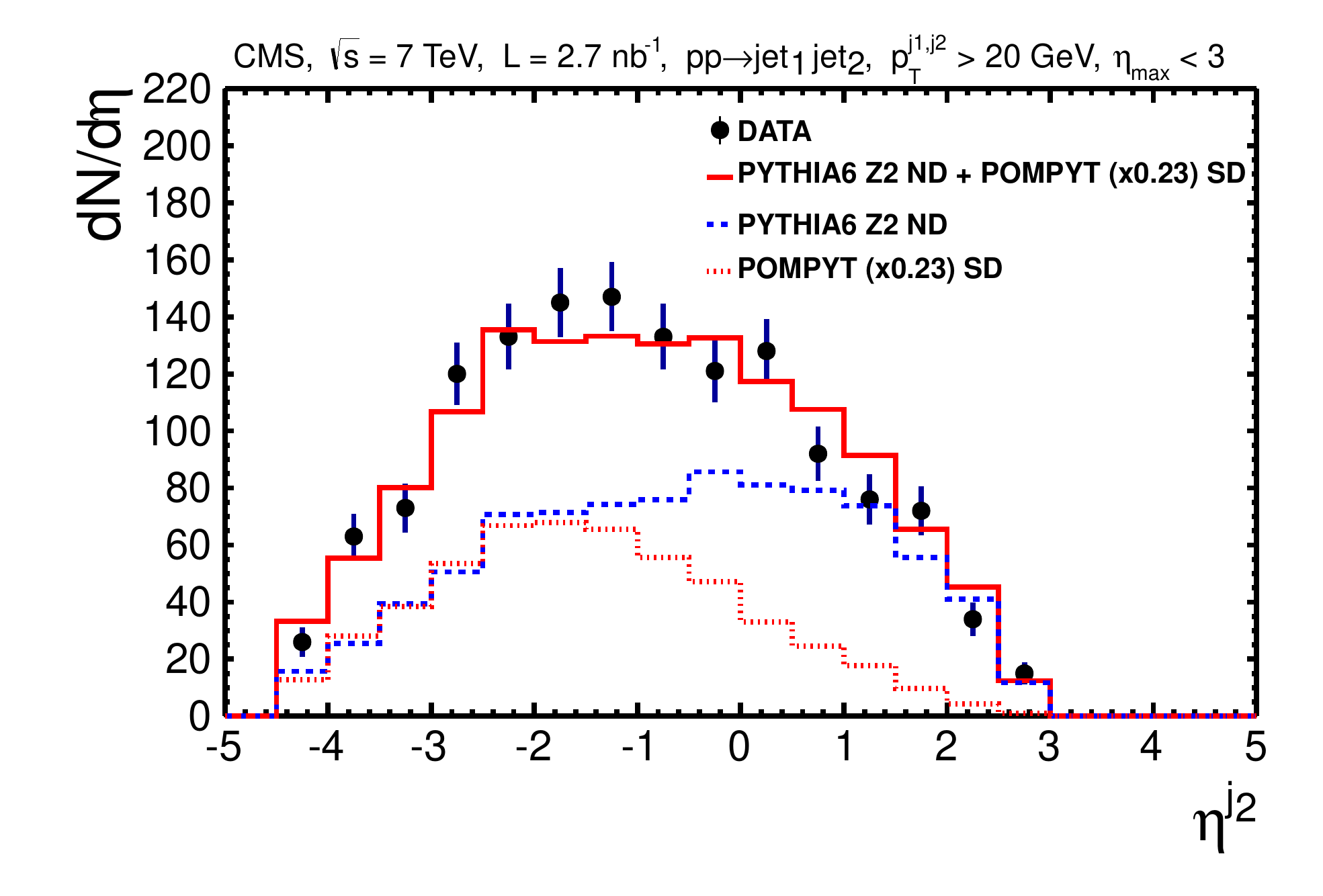}
    \caption{
Reconstructed pseudorapidity distributions of the leading (top) and second-leading
(bottom) jets after the $\eta_\text{max}<3$ selection
(black dots) compared to three detector-level MC simulations (histograms).
Events with the  $\eta_\text{min}>-3$ condition are also included in
the figure with $\eta^\mathrm{j1,j2} \rightarrow -\eta^\mathrm{j1,j2}$.
The error bars indicate the statistical
uncertainty. The predictions of the non-diffractive (\PYTHIA{}6 Z2) and diffractive (\textsc{pompyt}, scaled by
the value quoted in the legend) contributions and their sum are also
shown.
The sum of the predictions of the two MC simulations is normalised
to the number of events in the corresponding distributions for the data.
}
    \label{fig:jet_eta_gap}
  \end{center}
\end{figure}

The pseudorapidity distributions of the two leading jets for events selected
with the $\eta_\text{max}<3$ requirement are presented in Fig.~\ref{fig:jet_eta_gap}.
Events with the $\eta_\text{min}>-3$ condition are also included in Fig.~\ref{fig:jet_eta_gap}
with $\eta^\mathrm{j1,j2} \rightarrow -\eta^\mathrm{j1,j2}$.
The pseudorapidity gap condition enhances the diffractive component in the data,
and selects events with the jets mainly in the hemisphere opposite to that of the gap.
A combination of \PYTHIA{}6 Z2 and \textsc{pompyt} events reproduces the data
reasonably well; the relative normalisation of the models is optimised
with the procedure described in Section~\ref{ssec:reconstructed}.

\subsection{Reconstructed \texorpdfstring{\wxi}{xi} distributions and determination of the relative \textsc{POMPYT} and \textsc{PYTHIA{}6} normalisation\label{ssec:reconstructed}}
\label{sec:rec_xi}
\begin{figure*}[hbtp]
  \begin{center}
    \includegraphics[width=0.9\textwidth]{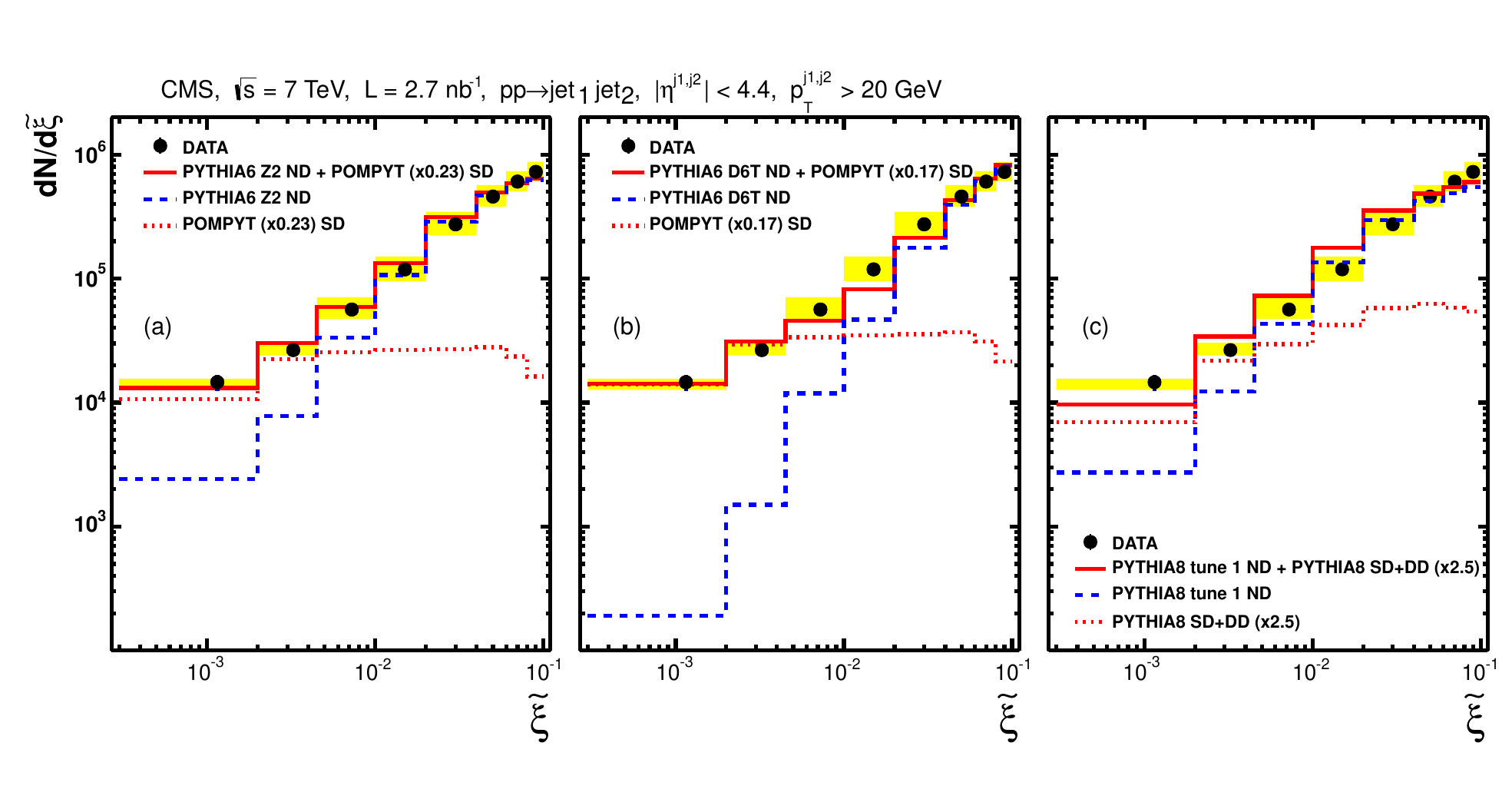}
    \caption{
Reconstructed \wxi  distribution compared to detector-level MC
predictions with and without diffractive dijet production.
The predictions of (a) \PYTHIA{}6 Z2 + \textsc{pompyt},
(b) \PYTHIA{}6 D6T + \textsc{pompyt}, and (c) \PYTHIA{}8 tune 1
are shown (in all the cases the relative diffractive contributions
from the MC simulation
are scaled by the values given in the legend).
The error bars indicate the statistical uncertainty, the band represents
the calorimeter energy scale uncertainty.
The sum of the predictions of the two MC simulations is normalised
to the number of events in the corresponding distributions for the data.
}
    \label{fig:xi_reco1}
  \end{center}
\end{figure*}

The reconstructed \wxi distribution is shown in Fig.~\ref{fig:xi_reco1}
before the $\eta_\text{max}$, $\eta_\text{min}$ selections. Here again, the shape of the
distribution can be described by the combination of diffractive and non-diffractive
MC models. The best combination was obtained by minimising the
difference between the \wxi distributions of the data and of
the sum of non-diffractive and diffractive models. The relative
contribution of diffractive dijets production and the overall normalisation of
the sum were found in this fit,
and the diffractive contribution was scaled accordingly.
The overall normalisation of the fit result is not relevant.
The effect of the calorimeter energy scale uncertainty, estimated
by varying by $\pm10\%$ the energy of all PF objects not associated with the leading jets,
is shown by the band.
The solid line in Fig.~\ref{fig:xi_reco1}(a) indicates the result of the fit,
according to which the diffractive dijet cross section predicted by  \textsc{pompyt}
should be multiplied by a factor ${\simeq}0.23$ to match the data.
The uncertainty of this correction factor was estimated by changing
the fitting procedure and was found to be ${\sim}20\%$.
Figure~\ref{fig:xi_reco1}(b)  presents the same data compared to
\PYTHIA{}6 D6T + \textsc{pompyt}; here the fit
requires the \textsc{pompyt}  normalisation to be scaled
by a factor of $\simeq$0.17.

Figure~\ref{fig:xi_reco1}(c) compares the data to \PYTHIA{}8 tune 1;
both the single-diffractive and the double-diffractive components are added
to the non-diffractive part, all simulated by \PYTHIA{}8. The result
of the fit is very different from that for \textsc{pomwig} and \PYTHIA{}6:
the normalisation of the diffractive components of \PYTHIA{}8 needs to be
multiplied by a factor ${\simeq}2.5$ to match the data.
This large difference is a consequence
of the different implementation of the pomeron flux in \PYTHIA{}8 and
\textsc{pompyt}.

In all three cases, after normalisation,
the shape of the reconstructed \wxi distribution
in the data is described satisfactorily by the MC models
(\PYTHIA{}6 Z2 + \textsc{pompyt} resulting in the best description).
However, the predicted non-diffractive component in the lowest \wxi
bin varies from about 0.1\% for \PYTHIA{}6 D6T to as much as 10--20\%
for \PYTHIA{}6 Z2 and \PYTHIA{}8.

\begin{figure}[hbtp]
  \begin{center}
    \includegraphics[width=\cmsFigWidth]{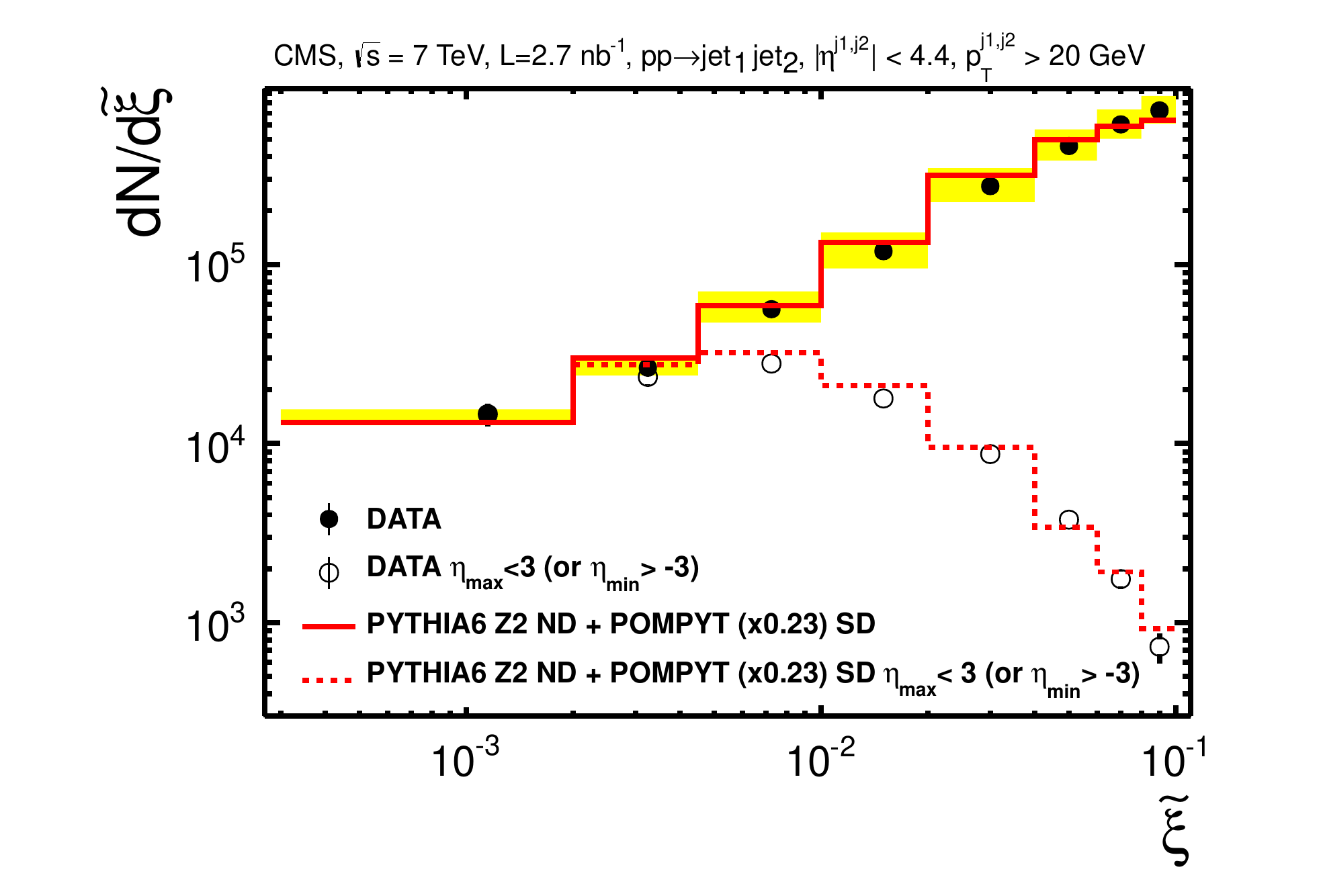}
    \caption{
Reconstructed \wxi distributions
with (open symbols) and without (closed symbols) the $\eta_\text{max}<3$ (or $\eta_\text{min}>-3$) condition
are compared to detector-level MC predictions including diffractive dijet production
(\PYTHIA{}6 Z2 + \textsc{pompyt}).
The error bars indicate the statistical uncertainty, the band represents
the calorimeter energy scale uncertainty.
The relative diffractive dijet contribution from the MC simulation has been
scaled by the factor 0.23.
The sum of the predictions of the two MC simulations is normalised
to the number of events in the corresponding distributions for the data.
}
    \label{fig:xi_reco2}
  \end{center}
\end{figure}

The effect of the $\eta_\text{max}<3$ ($\eta_\text{min}>-3$) requirement is illustrated
in Fig.~\ref{fig:xi_reco2}, where
the reconstructed \wxi distributions
with and without the  $\eta_\text{max}<3$ ($\eta_\text{min}>-3$) condition
are compared to MC simulations.
These pseudorapidity gap selections reject events at
high values of \wxi. The region of low
\wxi, where the diffractive
contribution dominates, is only marginally affected.
The data and MC simulations are in agreement at low \wxi.
The relative normalisation of \PYTHIA{}6 Z2 and \textsc{pompyt} in the
figure is the same as in Fig.~\ref{fig:xi_reco1}.

\section{Cross section determination and systematic uncertainties\label{sec:xsec}}
The differential cross section for dijet production  as a function  of \wxi is evaluated as

\begin{equation}
\frac{\rd\sigma_{jj}}{\rd\wxi} = \frac{N_{jj}^{i}}{L \cdot \epsilon \cdot A^{i}  \cdot\Delta\wxi^{i}}  \;\; ,
\end{equation}

where  $N_{jj}^{i}$ is the measured number of dijet events in the $i$-th
\wxi bin, $A^{i}$ is the correction factor defined as
the number of reconstructed MC events in that bin divided by
the number of generated events in the same bin,
$\Delta \wxi^{i}$ is the bin width, $L$ is the integrated
luminosity and $\epsilon$ is the trigger efficiency.
The factors $A^{i}$ include the effects of the geometrical acceptance of the
apparatus, that of all the selections listed in Section~\ref{sec:selection},
as well as the unfolding corrections to account for the finite
resolution of the reconstructed variables used in the analysis.
Various unfolding techniques
(bin-by-bin, SVD~\cite{svd} and Bayesian~\cite{bayes}) yield consistent
results and the bin-by-bin correction was kept.
In addition, the measured number of events, $N_{jj}^{i}$,
is corrected for the effect of pile-up.
This correction takes into account the probability of single pp interactions, evaluated on a run-by-run
basis, as well as the probability that pile-up interactions do not destroy the visible gap,
estimated with the minimum-bias MC samples (\PYTHIA{}6 Z1 and
\PYTHIA{}8 tune 1); the average correction is 1.07.
The cross section is measured for dijets with the axes in the
pseudorapidity range $|\eta^\mathrm{j1,j2}|<4.4$ and $p_{\mathrm{T}}^\mathrm{j1,j2}>20$\GeV
in the \wxi bins $0.0003<\wxi<0.002$,
$0.002<\wxi<0.0045$, and $0.0045<\wxi<0.01$.
The cross section results for $\wxip$ and
$\wxim$ are averaged, yielding the cross section as a function
of \wxi.

The systematic uncertainties are estimated by varying the selection criteria
and by modifying the analysis procedure as follows:

\begin{enumerate}

\item
The uncertainty on the jet energy scale varies between
$2\%$ and $9\%$ depending on the jet \pt and $\eta$~\cite{jet_correction}.
It decreases with the jet \pt and is typically higher at high $\eta$.
The energy of the reconstructed jets is varied accordingly.

\item
The effect of the uncertainty on the jet energy resolution is studied
by changing the resolution by up to $\pm10\%$ in the central region
($|\eta|<2.3$) and by up to $\pm20\%$ in the forward regions
($|\eta|>2.3$)~\cite{jet_correction}.

\item
The systematic uncertainty related to the \wxi
reconstruction is determined as follows:
(i) the effect of the calorimeter energy scale uncertainty is
estimated by varying the energy of all PF objects not associated with
the leading jets by ${\pm}10\%$;
(ii) the \pt  threshold for tracks is increased from 200 to 250\MeV;
(iii) the correction factor $C$ is varied by ${\pm}3\%$, i.e.
by its uncertainty (as discussed in Section~\ref{sec:selection}).

\item
The uncertainty on the correction factor $A^i$ in Eq.~(4) is estimated by
changing the MC models used to evaluate it. In addition, the relative fraction
of diffraction is changed by ${\pm}20\%$, \ie by the uncertainty of the
scaling factors obtained in the fits discussed in Section~\ref{ssec:reconstructed}.

\item
The sensitivity to pile-up is studied by restricting the analysis to events
with only one reconstructed vertex.

\item
The sensitivity to the jet reconstruction procedure is studied by repeating
the analysis with jets reconstructed only with calorimetric information instead of
particle-flow objects.  This affects the results by 4\% at most.

\item
The difference in the results obtained for the cross section as a function
of $\wxip$  and  $\wxim$ is found to be
less than $11\%$ and is included in the systematic uncertainty.

\item
The uncertainty on the trigger efficiency is estimated from the comparison of
the turn-on curves as a function of the jet \pt in the minimum-bias data
and the MC simulation. The resulting uncertainty is 3\%.

\item
The uncertainty on the integrated luminosity is estimated to be
$4\%$~\cite{lumi1,lumi2}.

\end{enumerate}

The total systematic uncertainty is calculated as the quadratic sum of the individual
contributions. The resulting uncertainty of the cross section measurement
is ${\sim}30\%$, dominated by the jet energy scale.
The effect of each systematic check on the cross section uncertainty
is given in Table~\ref{tab:syst}.

\begin{table*}[htbH]
\renewcommand{\arraystretch}{1.4}
\begin{center}
\topcaption{
Contributions to the systematic uncertainty on the dijet cross section
in the three lowest \wxi bins considered.
The total systematic uncertainty calculated as the quadratic sum of
the individual contributions is given in the last row.
\label{tab:page_layout}}
\scotchrule[llll]
Uncertainty source  & $0.0003<\wxi<0.002$ & $0.002<\wxi<0.0045$ & $0.0045<\wxi<0.01$  \\
\hline
1. Jet energy scale & $({+26};{-19})\%$   & $({+21};{-20})\%$ & $({+28};{-16})\%$  \\
2. Jet energy resolution & $({+6};{-4})\%$   & $({+4};{-3})\%$ & $({+3};{-2})\%$  \\
3. PF energy, \pt threshold, C &  $({+7};{-15})\%$   &  $({+14};{-8})\%$ &  $({+12};{-11})\%$  \\
4. MC model uncertainty &  $({+5};{-3})\%$   &  $({+2};{-14})\%$ &  $({+3};{-1})\%$  \\
5. One-vertex selection  &  $({+6};{-0})\%$   &  $({+0};{-1})\%$ &  $({+1};{-0})\%$  \\
6. Jet objects (Calorimeter, PF)  &  $({+0};{-4})\%$   &  $({+0};{-4})\%$ &  $({+2};{-4})\%$  \\
7. $\wxip$, $\wxim$ difference   &  $\pm8\%$   &  $\pm8\%$ &  $\pm11\%$  \\
8. Trigger efficiency &  $\pm3\%$   &  $\pm3\%$ &  $\pm3\%$  \\
9. Luminosity &  $\pm4\%$   &  $\pm4\%$ &  $\pm4\%$  \\
\hline
Total error &  $({+30};{-26})\%$   &  $({+27};{-29})\%$ &  $({+33};{-23})\%$  \\
\donescotchrule
\label{tab:syst}
\end{center}
\end{table*}

\section{Results\label{sec:results}}

\begin{figure}[hbtp]
  \begin{center}
    \includegraphics[width=\cmsFigWidth]{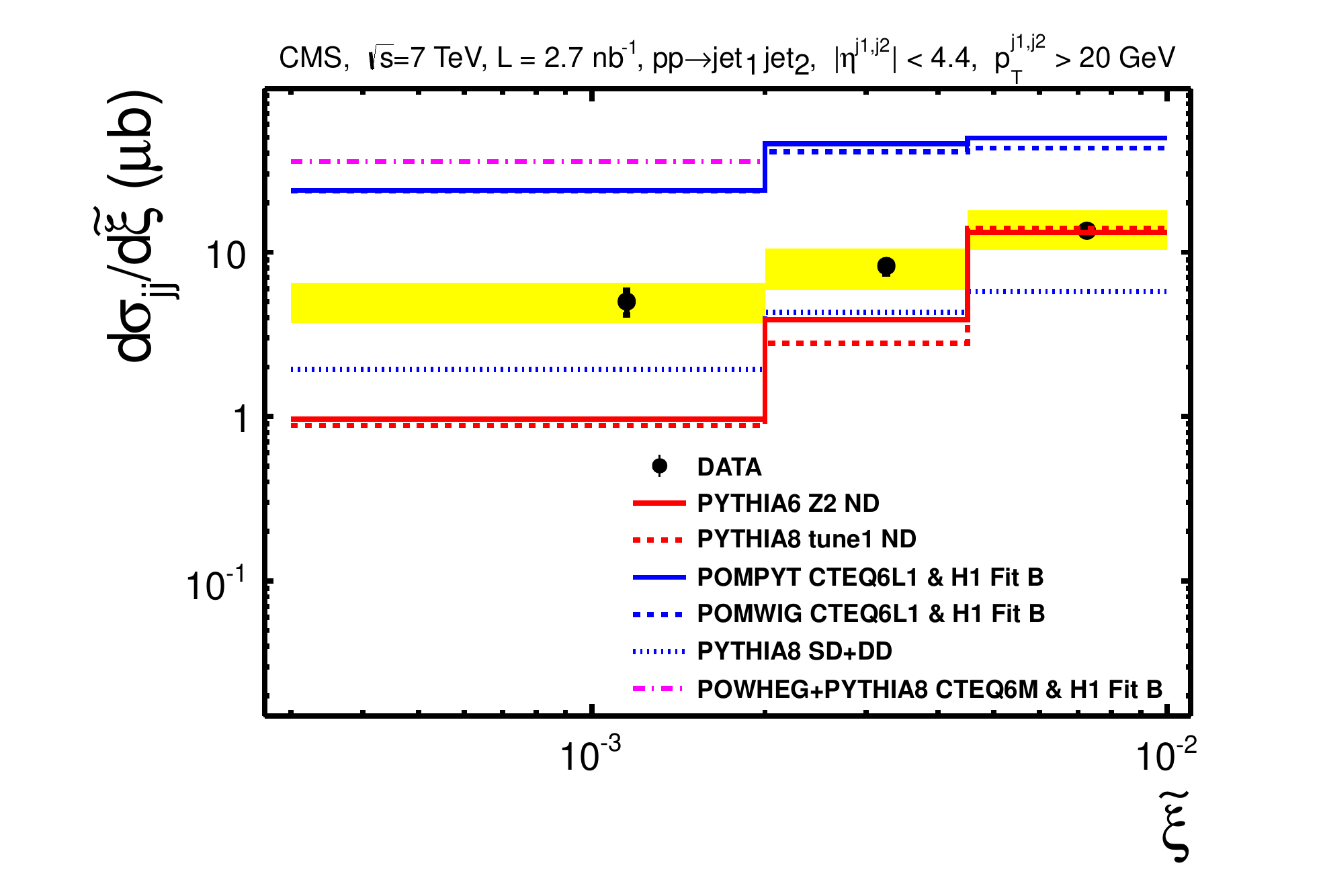}
    \caption{
The differential cross section for inclusive dijet production as a function
of \wxi
for jets with axes in the range $|\eta^\mathrm{j1,j2}|<4.4$  and
$p_{\mathrm{T}}^\mathrm{j1,j2}>20$\GeV.
The points are plotted at the centre of the bins.
The error bars indicate the statistical uncertainty and the band
represents the systematic uncertainties added in
quadrature. The predictions of non-diffractive (\PYTHIA{}6 Z2 and
\PYTHIA{}8 tune 1) and diffractive (\textsc{pompyt} SD, \textsc{pomwig} SD
and \PYTHIA{}8 SD+DD) MC generators are also shown, along with that of the
NLO calculation based on \POWHEG (first bin only).
The predictions of \textsc{pompyt} and \textsc{pomwig}  in the first \wxi bin are identical.
}
    \label{fig:xsec1}
  \end{center}
\end{figure}

Table~\ref{tab:xsec} and Fig.~\ref{fig:xsec1}  present the differential cross
section for dijet production as a function of \wxi.
The data are compared to the predictions of non-diffractive (\PYTHIA{}6 Z2
and \PYTHIA{}8 tune 1) and diffractive
(\textsc{pompyt} SD, \textsc{pomwig} SD, \PYTHIA{}8 SD+DD, and \POWHEG)
models.
The normalisation of the predictions is absolute,
unlike in Fig.~\ref{fig:xi_reco1}.

\begin{table}[htbH]
\renewcommand{\arraystretch}{1.4}
\begin{center}
\topcaption{
Differential cross section for inclusive dijet production as a function
of \wxi for jets with $p_{\mathrm{T}}^\mathrm{j1,j2}>20$\GeV
and jet-axes in the pseudorapidity range $|\eta^\mathrm{j1,j2}|<4.4$.
}
\scotchrule[ll]
\wxi bin & $\rd\sigma_{jj}/\rd\wxi$ ($\mu$b)    \\
\hline
$0.0003<\wxi<0.002$ & $5.0\pm0.9\stat\,  ^{+1.5}_{-1.3}\syst$   \\
$0.002<\wxi<0.0045$ & $8.2\pm0.9\stat\, ^{+2.2}_{-2.4}\syst$   \\
$0.0045<\wxi<0.01$  & $13.5\pm0.9\stat\, ^{+4.5}_{-3.1}\syst$  \\
\donescotchrule
\label{tab:xsec}
\end{center}
\end{table}

The following conclusions can be drawn from Fig.~\ref{fig:xsec1}:

\begin{itemize}

\item
The generators \PYTHIA{}6 Z2 and \PYTHIA{}8 tune 1, without
hard diffraction, cannot by themselves describe the low-\wxi
data, especially in the first bin, $0.0003<\wxi<0.002$.

\item
It was noted already in Section~\ref{ssec:reconstructed} that the contribution of SD MC models,  e.g.
\textsc{pomwig} and \textsc{pompyt}, is needed to describe the low-\wxi
data, reflecting the
presence of hard diffractive events in this region. However, these MC models predict
more events than are observed, by a factor of about 5 in the lowest \wxi bin.

\item
The ratio of the measured cross section to that expected from the \textsc{pompyt} and
\textsc{pomwig} simulations is $0.21 \pm 0.07$
in the first \wxi bin, where the non-diffractive contribution is small.
This ratio can be taken as an upper limit of the rapidity gap
survival probability (not simulated by the event generators considered).
This is an upper limit because
the measured cross section includes a
contribution from proton-dissociative events in which the scattered proton
is excited into
a low mass state, which escapes undetected in the forward
region; the dPDFs also include a proton-dissociative
contribution. If the amount of proton-dissociative events in the data is
assumed to be 41\%, as estimated at particle level with \PYTHIA{}8, and that in
the dPDFs is taken to be 23\%~\cite{hera_diffraction1}, then this upper
limit can be turned into an estimate of the rapidity gap survival
probability of $0.12 \pm 0.05$.

\item
\textsc{pompyt} and \textsc{pomwig} are leading-order (LO) MC generators. If \POWHEG
is used to predict the diffractive cross section at NLO in the first \wxi
bin and \PYTHIA{}8 tune 1 is used for hadronisation, the ratio between data and
predictions becomes $0.14 \pm 0.05$.
With the assumptions just discussed on the proton-dissociative
contribution, the rapidity gap survival probability becomes $0.08 \pm 0.04$.

\item
Figure~\ref{fig:xsec1} also shows that the normalisation of the SD+DD \PYTHIA{}8
prediction disagrees with that of \textsc{pompyt} and \textsc{pomwig}, and would have
to be scaled up by a factor about two to match the data. This is a consequence of the
different modelling of diffraction in these generators: while they all use the same
H1 dPDFs, the parameterisation of the pomeron flux in \PYTHIA{}8 is
different -- and notably not that used in the H1 fit. Because of this,
\PYTHIA{}8 (version 8.135) cannot be used to extract the rapidity gap
survival probability.

\end{itemize}

While the rapidity gap survival probability measured at the
Tevatron~\cite{cdf_dijet,newcdf}
is close to that found in the present analysis, the two measurements
cannot be directly
compared because of the different kinematic regions they cover: $0.035<\xi<0.095$
for the CDF data and $0.0003<\wxi<0.002$
for the present CMS data. This difference is relevant because the
rapidity gap survival
probability depends on the parton momentum $x$ and is expected to increase with
decreasing $x$ (and hence $\xi$): from about 0.05 at $x=10^{-1}$ to about 0.3
for $x=10^{-3}$ according to Ref.~\cite{klasen}.

\section{Summary\label{sec:summary}}
The differential cross section for dijet production as a function of
\wxi,
a variable that approximates the fractional momentum loss of the proton
in single-diffractive processes, has been measured with the CMS detector
for events with at least two jets with $p_{\mathrm{T}}^\mathrm{j1,j2}>20$\GeV in the pseudorapidity
region $|\eta^\mathrm{j1,j2}|<4.4$. The results are compared to diffractive
(\textsc{pompyt},
\textsc{pomwig}, and \PYTHIA{}8 SD+DD) and non-diffractive
(\PYTHIA{}6 Z2, D6T, and \PYTHIA{}8 tune 1) MC models.
The low-\wxi data show a significant contribution from diffractive dijet production,
observed for the first time at the LHC. The associated rapidity gap survival probability is
estimated.
Leading-order diffractive  generators (\textsc{pompyt} and \textsc{pomwig}),
based on dPDFs
from the HERA experiments, overestimate the measured  cross section and
their normalisation needs to be scaled down by a factor of $\sim$5. This
factor can be interpreted as the effect of the rapidity gap survival probability.
The results are also compared with  NLO predictions.
The rapidity gap survival probability, estimated from the comparison of the cross
section measured in the first bin, $0.0003<\wxi<0.002$,
with LO and NLO diffractive MC models,
ranges from $0.08 \pm 0.04$ (NLO) to $0.12 \pm 0.05$ (LO).

\section*{Acknowledgements}
We would like to thank M.~Diehl, M.G.~Ryskin and L.~Trentadue  for their valuable suggestions.
We congratulate our colleagues in the CERN accelerator departments for
the excellent performance of the LHC machine.
We thank the technical and administrative staff at CERN and other CMS
institutes, and acknowledge support from BMWF and FWF (Austria); FNRS
and FWO (Belgium); CNPq, CAPES, FAPERJ, and FAPESP (Brazil); MES
(Bulgaria); CERN; CAS, MoST, and NSFC (China);  COLCIENCIAS
(Colombia); MSES (Croatia); RPF (Cyprus); MoER, SF0690030s09 and ERDF
(Estonia); Academy of Finland, MEC, and HIP (Finland); CEA and
CNRS/IN2P3 (France); BMBF, DFG, and HGF (Germany); GSRT (Greece); OTKA
and NKTH (Hungary); DAE and DST (India); IPM (Iran); SFI (Ireland);
INFN (Italy); NRF and WCU (Korea); LAS (Lithuania); CINVESTAV,
CONACYT, SEP, and UASLP-FAI (Mexico); MSI (New Zealand); PAEC
(Pakistan); MSHE and NSC (Poland); FCT (Portugal); JINR (Armenia,
Belarus, Georgia, Ukraine, Uzbekistan); MON, RosAtom, RAS and RFBR
(Russia); MSTD (Serbia); SEIDI and CPAN (Spain); Swiss Funding
Agencies (Switzerland); NSC (Taipei); ThEP, IPST and NECTEC
(Thailand);  TUBITAK and TAEK (Turkey); NASU (Ukraine); STFC (United Kingdom); DOE and NSF (USA).
Individuals have received support from the Marie-Curie programme and
the European Research Council (European Union); the Leventis
Foundation; the A. P. Sloan Foundation; the Alexander von Humboldt
Foundation; the Austrian Science Fund (FWF); the Belgian Federal
Science Policy Office; the Fonds pour la Formation \`a la Recherche
dans l'Industrie et dans l'Agriculture (FRIA-Belgium); the Agentschap
voor Innovatie door Wetenschap en Technologie (IWT-Belgium);
the Ministry of Education, Youth and Sports (MEYS) of Czech Republic;
the Council of Science and Industrial Research, India; the Compagnia
di San Paolo (Torino); and the HOMING PLUS programme of Foundation for
Polish Science, cofinanced from European Union, Regional Development Fund.

\bibliography{auto_generated}   

\cleardoublepage \appendix\section{The CMS Collaboration \label{app:collab}}\begin{sloppypar}\hyphenpenalty=5000\widowpenalty=500\clubpenalty=5000\textbf{Yerevan Physics Institute,  Yerevan,  Armenia}\\*[0pt]
S.~Chatrchyan, V.~Khachatryan, A.M.~Sirunyan, A.~Tumasyan
\vskip\cmsinstskip
\textbf{Institut f\"{u}r Hochenergiephysik der OeAW,  Wien,  Austria}\\*[0pt]
W.~Adam, E.~Aguilo, T.~Bergauer, M.~Dragicevic, J.~Er\"{o}, C.~Fabjan\cmsAuthorMark{1}, M.~Friedl, R.~Fr\"{u}hwirth\cmsAuthorMark{1}, V.M.~Ghete, J.~Hammer, N.~H\"{o}rmann, J.~Hrubec, M.~Jeitler\cmsAuthorMark{1}, W.~Kiesenhofer, V.~Kn\"{u}nz, M.~Krammer\cmsAuthorMark{1}, I.~Kr\"{a}tschmer, D.~Liko, I.~Mikulec, M.~Pernicka$^{\textrm{\dag}}$, B.~Rahbaran, C.~Rohringer, H.~Rohringer, R.~Sch\"{o}fbeck, J.~Strauss, A.~Taurok, W.~Waltenberger, G.~Walzel, E.~Widl, C.-E.~Wulz\cmsAuthorMark{1}
% ...
\vskip\cmsinstskip
\dag:~Deceased\\
1:~~Also at Vienna University of Technology, Vienna, Austria\\
% ...

\end{sloppypar}
\end{document}